\documentclass[twocolumn,
superscriptaddress,
showpacs,
 amsmath,amssymb,
 aps,
 prl,
floatfix,
]{revtex4}

\usepackage{graphicx}
\usepackage{dcolumn}
\usepackage{bm}

\usepackage[T1]{fontenc}
\usepackage{lmodern}

\begin{document}

\preprint{APS/123-QED}

\title{Polariton Condensation in a One-Dimensional Disordered Potential}

\author{F.~Manni}%
 \email{francesco.manni@epfl.ch}
\author{K.~G.~Lagoudakis}%
\affiliation{%
  Institute of Condensed Matter Physics, \'{E}cole Polytechnique F\'ed\'erale de Lausanne (EPFL), CH-1015 Lausanne, Switzerland}%
\author{B.~Pietka}%
\affiliation{%
  Institute of Experimental Physics, University of Warsaw, ul. Hoza 69, 00-681 Warszawa, Poland}%
\author{L.~Fontanesi}%
\affiliation{%
  Institute of Theoretical Physics, \'{E}cole Polytechnique F\'ed\'erale de Lausanne (EPFL), CH-1015 Lausanne, Switzerland}%

\author{M.~Wouters}%
\affiliation{TQC, Departement Fysica, Universiteit Antwerpen, Belgi\"e}

\author{V.~Savona}%
\affiliation{Institute of Theoretical Physics, \'{E}cole Polytechnique F\'ed\'erale de Lausanne (EPFL), CH-1015 Lausanne, Switzerland}

\author{R.~Andr\'e}%
\affiliation{%
  Institut N\'eel, CNRS, Grenoble, France\\}%

\author{B.~Deveaud-Pl\'edran}%
\affiliation{%
  Institute of Condensed Matter Physics, \'{E}cole Polytechnique F\'ed\'erale de Lausanne (EPFL), CH-1015 Lausanne, Switzerland}%

\date{\today}

\begin{abstract}
We study the coherence and density modulation of a non-equilibrium exciton-polariton condensate in a one-dimensional valley with disorder. By means of interferometric measurements we evidence a modulation of the first-order coherence function and we relate it to a disorder-induced modulation of the condensate density, that increases as the pump power is increased. The non-monotonous spatial coherence function is found to be the result of the strong non-equilibrium character of the one-dimensional system, in the presence of disorder.
\end{abstract}

\maketitle


Quantum degenerate Bose gases in external potentials display a variety of behaviors in a large number of systems, ranging from $^4$He in porous media \cite{reppy_superfluid_1992}, ultracold atoms in magnetic traps, optical lattices \cite{stringari_bec} and speckle potentials \cite{sanchez-palencia_disordered_2010} to exciton-polaritons in semiconductor microcavities.
In atomic systems, the transition from the superfluid to the Mott insulator was observed under the application of a periodic potential (optical lattice) \cite{greiner_quantum_2002}. More recently, evidence for a Bose glass phase was obtained for a Bose gas in a random potential \cite{deissler_delocalization_2010,pasienski_disordered_2010}. The Bose Glass (BG) phase is characterized by a vanishing superfluid fraction and an exponential decay of the spatial coherence and, in contrast to the Mott insulating state, it is compressible \cite{fisher_boson_1989}. Because the effect of disorder on a quantum system is more pronounced in lower dimensionality, the studies on the quantum phases of ultracold atomic gases focused on the special case of one dimensional (1D) quantum gases \cite{deissler_delocalization_2010,chen_phase_2008,roati_anderson_2008,billy_direct_2008}.

Since the convincing demonstration of polariton BEC by several groups \cite{kasprzak_bose-einstein_2006,balili_bose-einstein_2007,deng_spatial_2007,wertz_spontaneous_2010}, microcavity polaritons have attracted much interest as a physical realization of the quantum degenerate Bose gas. Exciton-polaritons are quasi-particles resulting from the strong coupling of a photon and an exciton. The half-light half-matter nature of polaritons confers them a very small effective mass ($10^{-4}$ times the mass of a free electron) coming from the photon component, while their excitonic part allows them to interact with each other. Both features are favorable to achieve condensation \cite{kasprzak_bose-einstein_2006}, which has been demonstrated even up to room temperature \cite{christmann_room_2008}, making of polaritons a model low-dimensional Bose-gas \cite{lagoudakis_quantized_2008,lagoudakis_observation_2009,amo_collective_2009,amo_superfluidity_2009}. However, it is of crucial importance to underline that, because of the exciton-polariton finite lifetime, their condensation is a non-equilibrium process, giving rise to a phenomenology which bears substantial differences with respect to the case of equilibrium BEC of atomic systems. Due to the structure of microcavities, the polariton gas is at most two-dimensional; by employing advanced growth techniques, one and zero dimensional systems can be manufactured as well \cite{wertz_spontaneous_2010,el_daif_polariton_2006}. Contrarily to ultracold atoms, disorder is present as a natural feature in microcavities, as a consequence of the fabrication process.
\begin{figure}[tb]
\includegraphics[width=0.4\textwidth]{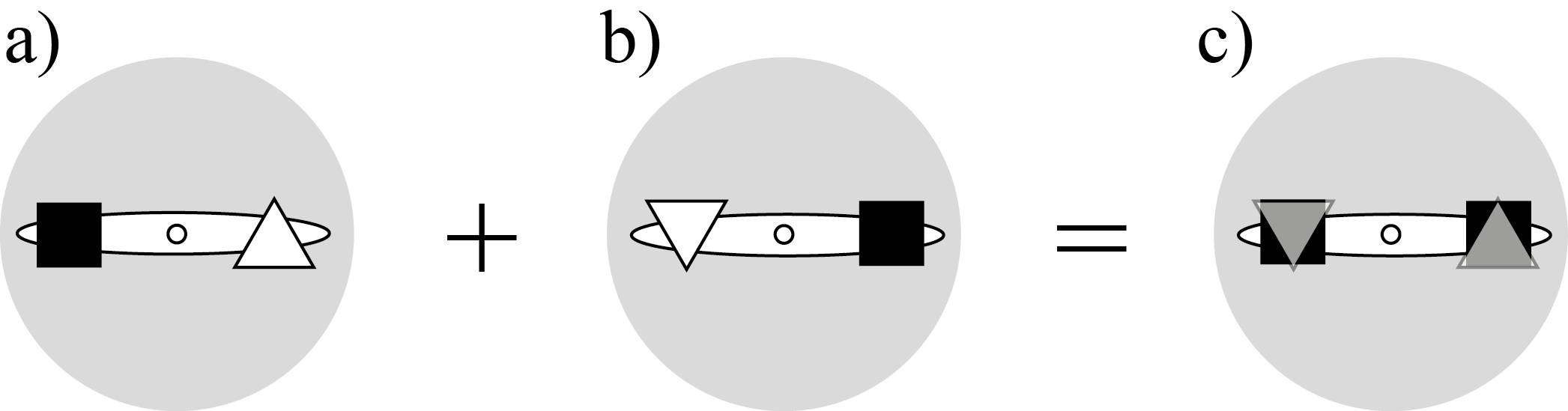}
\caption{\label{fig:michel_int} Michelson interferometer in the mirror-retroreflector configuration: a) the mirror arm is superposed to b) the retroreflector arm, which is a centrosymmetric replica of it, in order to get the c) interferogram. This scheme allows to probe the coherence of points of the condensate that are spatially separated from each other. The central point, which is overlapped with itself, is called the autocorrelation point}
\end{figure}

Recently, the condensation of 1D polaritons was achieved by Wertz et al. \cite{wertz_spontaneous_2010} in a GaAs polariton wire. They observed the spontaneous formation of spatial coherence under tightly focused non-resonant excitation. Due to the high quality of their GaAs microcavity and to the ballistic expansion of the polariton gas under their experimental conditions, the disorder played a rather minor role in these experiments. Although difficult to understand at first sight, these results have been reproduced by theory after taking into account the non-equilibrium character of the polariton system and some novel relaxation processes \cite{wouters_energy_2010}.

On the contrary, the CdTe sample employed here allows us to investigate the spatial coherence of a polariton condensate in a strong disorder. This natural feature seems promising in order to observe the BG to superfluid phase transition. However, as will be shown further on, we did not find any evidence of BG but rather a direct transition from a non-coherent gas of polaritons to the condensed state is observed. The transition to the condensed state is accompanied by a particular modulation of the first-order spatial coherence function that becomes more and more pronounced as the density of particles in the system is increased. We show how this feature is related to the observed disorder-induced modulation of the 1D condensate density. The particular non-monotonous behavior of the spatial coherence function that we find confirms the strong non-equilibrium character of our system in presence of disorder.

Our system consists of a CdTe/CdMgTe semiconductor microcavity with 16 quantum-wells, with an overall vacuum Rabi splitting of 26 meV. The main feature of our CdTe sample is the presence of a strong photonic disorder, which appears in various configurations at different positions on the sample. The presence of disorder affects the polariton condensation phenomenology, giving rise to density modulations.

We focused our attention to well defined one-dimensional shaped disorder valleys in the sample, which are delimited at the sides by high energy barriers, determining the linear shape of the polariton condensate. Furthermore, we made sure to find a region in the valley where condensation occurs in a single state, contrary to the generic behavior in our sample, where condensation usually occurs in multiple spatially overlapping energy states \cite{krizhanovskii_coexisting_2009}. Therefore, on such a particular location in the sample, it becomes possible to isolate and study the physics of a single polariton condensate, without the complications deriving from the mode competition between multiple condensates. The interest of the study of such a one-dimensional state lies in the fact that the effects of disorder are more pronounced in the 1D topology.

\begin{figure}[tb]
\includegraphics[width=0.48\textwidth]{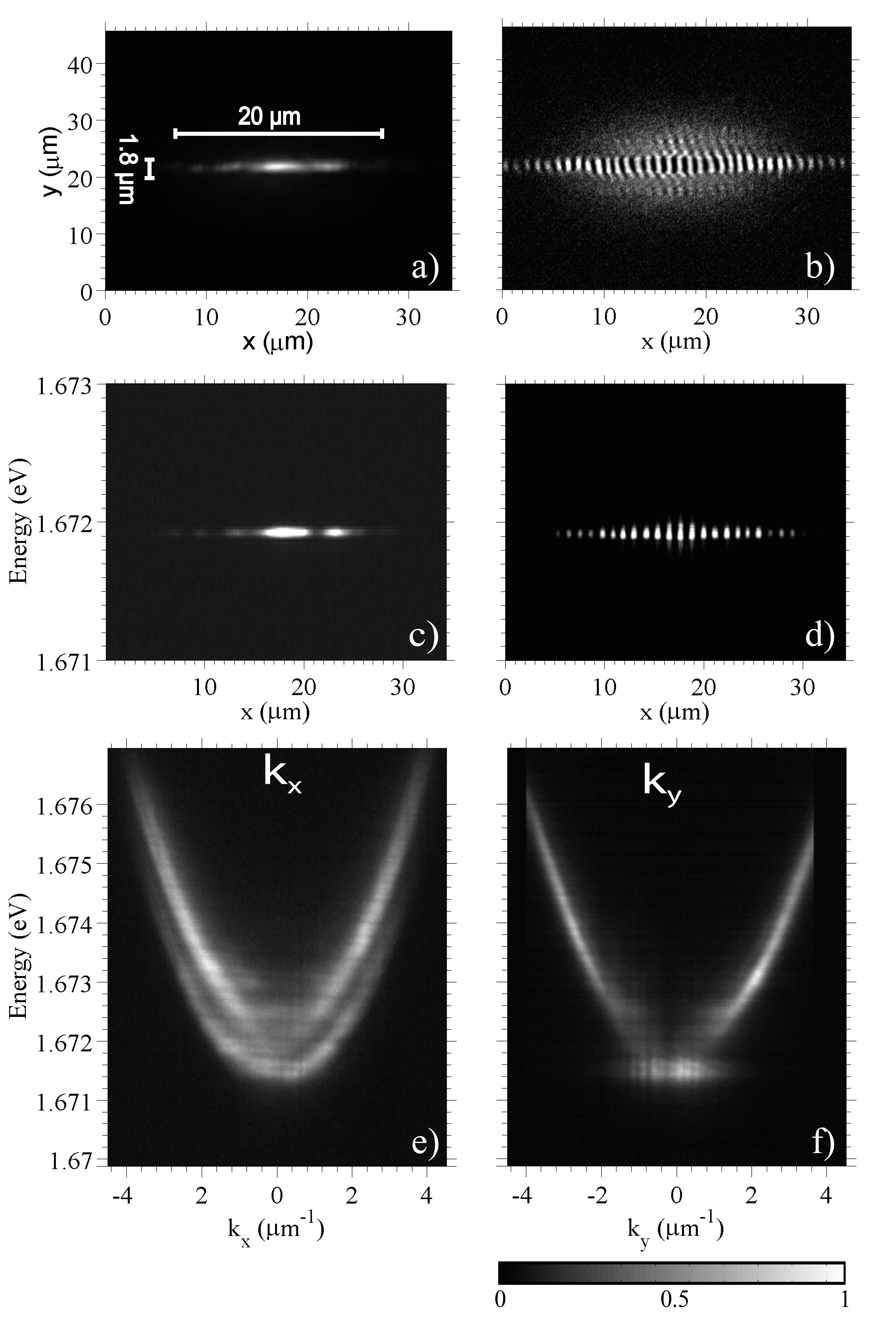}
\caption{\label{fig:figure1} a) Real space photoluminescence coming from the surface of the sample at a specific 1D disorder valley and b) associated interference pattern (both taken above condensation threshold). It can be seen that the coherence manifests only along the linear disorder valley. The corresponding real space energy spectrum are shown in c) and d) respectively, where the energy line of a single 1D state can be observed. The polariton dispersion, below condensation threshold, is shown in e) along $k_{x}$ (along the linear condensate) and in f) along $k_{y}$ (1D confinement direction). Note: the colorscale of (b), (c) and (d) is saturated to make the signal modulation more visible.}
\end{figure}

In our experimental setup, the sample is kept close to liquid Helium temperature ($\approx 10$ K). We excite the sample non-resonantly using a Ti:Sapphire monomode laser with a quasi-continuous gaussian beam. The excitation beam is focused on the surface of the sample with a high numerical aperture microscope objective, with an illumination spot diameter of approximately 20 $\mu$m. The photoluminescence (PL) signal is sent to a Michelson interferometer in the mirror-retroreflector configuration, that allows to measure in a direct way the spatial coherence of our condensate, quantified by $g^{(1)}(x_{1},x_{2}) = \frac{\langle\hat{\Psi}^{\dagger}(x_{1})\hat{\Psi}(x_{2})\rangle} {\sqrt{\langle|\hat{\Psi}(x_{1})|^{2}\rangle\langle|\hat{\Psi}(x_{2})|^{2}\rangle}}$, where $\hat{\Psi}(x)$ is the polariton annihilation operator \cite{stringari_bec}. The retroreflected image is a centrosymmetric replica of the mirror arm image. The interferometric superposition of the two arms makes points, symmetric with respect to the so-called autocorrelation point (the center of the image) to overlap in the interferogram (as sketched in Fig.\ref{fig:michel_int}). This allows to probe the spatial coherence between points of the condensate, that are spatially separated from each other, through the analysis of the interference fringes \cite{kasprzak_bose-einstein_2006}. The interferometric measurement gives direct access to the first-order spatial coherence function $g^{(1)}(r) = g^{(1)}(\frac{r}{2},-\frac{r}{2})$, where $r$ is the distance to the autocorrelation point along the 1D condensate.

In order to achieve condensation, we increase the pump power above a certain threshold power, $P_{th} = 100 \mu$W, injecting in the system a polariton population which is sufficient for stimulated scattering towards the ground state to occur and macroscopically populate it. The real-space luminescence, coming from the one-dimensional disorder valley studied in this work, is shown in Fig.\ref{fig:figure1}.a. In Fig.\ref{fig:figure1}.b one can see the interference fringes that occur at the output of the modified Michelson interferometer along the whole length of the one dimensional condensate. The corresponding spectra are shown in Fig.\ref{fig:figure1}.c and Fig.\ref{fig:figure1}.d respectively, where the presence of a single energy state is confirmed. Further confirmation comes from a tomographic reconstruction (not shown) displaying the spectrally resolved emission in the whole k-space. Below the condensation threshold for polaritons, a single 1D confined state is observed, which shows a parabolic dispersion only along one of the two orthogonal k-space axes, coexisting with the whole parabola of the extended 2D state. More specifically in Fig.\ref{fig:figure1}.e we show the polariton dispersion along the $k_{x}$ direction, which corresponds to the linear condensate direction, and along $k_{y}$ (Fig.\ref{fig:figure1}.f), the one-dimensional confinement direction (both images are recorded below condensation threshold). It can be seen that the lowest state at $\approx 1.6714$ eV is 1D confined, featuring a parabolic dispersion only along the linear condensate direction. When we increase the excitation power above $P_{th}$, condensation occurs in the 1D confined state whilst the 2D extended states are still far below threshold. The emission from the 2D polariton states is thus not visible anymore with respect to the 1D condensate emission. It can be also noticed that, above threshold, the 1D confined state is blueshifted by $\approx 0.5$ meV (see Fig. \ref{fig:figure1}.f versus \ref{fig:figure1}.c) due to the interactions of the increased polariton population with the excitonic reservoir. This value is typical and in line with our previous experimental observations.

We performed both real-space photoluminescence and coherence measurements over a wide range of excitation powers, starting from below the condensation threshold up to almost three times the threshold power, a range over which we are able to track the single 1D confined state and study its intrinsic properties. For higher pump powers we observe the occurrence of condensation in multiple states and mode competition, which is due to the non-equilibrium nature of polariton condensation.

\begin{figure}[tb]
\includegraphics[width=0.48\textwidth]{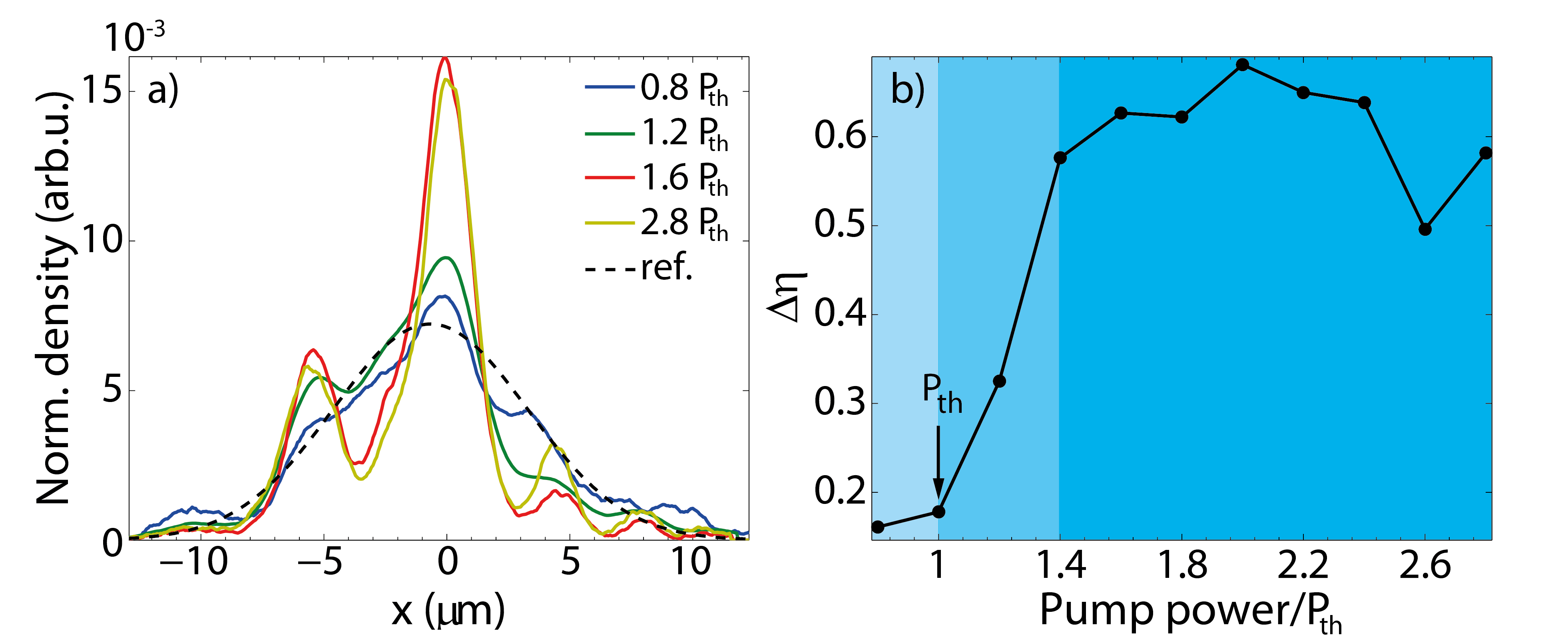}
\caption{\label{fig:figure3} a) Real space normalized polariton density profiles along the 1D condensate, for different values of the excitation power. The black dotted curve is the reference profile considered in the analysis of the disorder-induced density modulations. b) Standard deviation of the measured density profiles with respect to the gaussian reference profile (background color gradient identifies three different regimes, as explained in the text).}
\end{figure}

The real space PL measurements allow us to directly access the polariton density, which is reported in Fig.\ref{fig:figure3}.a for different values of the pump power. Below threshold, the classical polariton gas is characterized by a rather smooth density, because the effective temperature of the thermal polaritons is larger than the average amplitude of the disorder potential. Above threshold, a single energy delocalized condensed state builds up, allowed by the polariton interactions that bring in phase different spatial regions of the condensate: a process known as mode synchronization \cite{baas_synchronized_2008}. Instead of a smooth density, the condensed phase features a more and more pronounced modulation of the polariton density as the excitation power is increased. The observed modulation is the result of localization effects in the disorder.

In order to point out the role of disorder in a more quantitative way, we carried out further analysis, in analogy with the usual analysis employed in atomic condensates \cite{clement_density_2008}. We considered a reference density distribution, representative of the polariton condensate density in an ideal disorder-less case \cite{ref_prof_details}. We assumed as a reference the gaussian profile, $\rho_{0}(x)$, shown in Fig.\ref{fig:figure3}.a with a dashed black line. The standard deviation, $\Delta\eta$, with respect to this reference is calculated for the measured density profiles at all different values of pump power, according to the formula $\Delta\eta = \sqrt{\frac{1}{N}\sum_{x}\left(\frac{\rho(x) - \rho_{0}(x)}{\rho_{0}(x)}\right)^{2}}$,
where $\rho(x)$ is the normalized measured density. The sum over $x$ extends over the N central points (corresponding to the interval from -8 $\mu m$ to 8 $\mu m$) along the linear condensate, excluding the tails, strongly affected by noise. In accordance with the qualitative observation, for increasing excitation power the standard deviation increases (Fig.\ref{fig:figure3}.b), confirming that the modulation due to the disorder becomes increasingly relevant. The background color gradient in Fig.\ref{fig:figure3}.b replicates the three regimes identified in the behavior with pump power of the first-order coherence function: non-coherent gas, coherence build-up and fully coherent polariton condensate.

\begin{figure}[tb]
\includegraphics[width=0.48\textwidth]{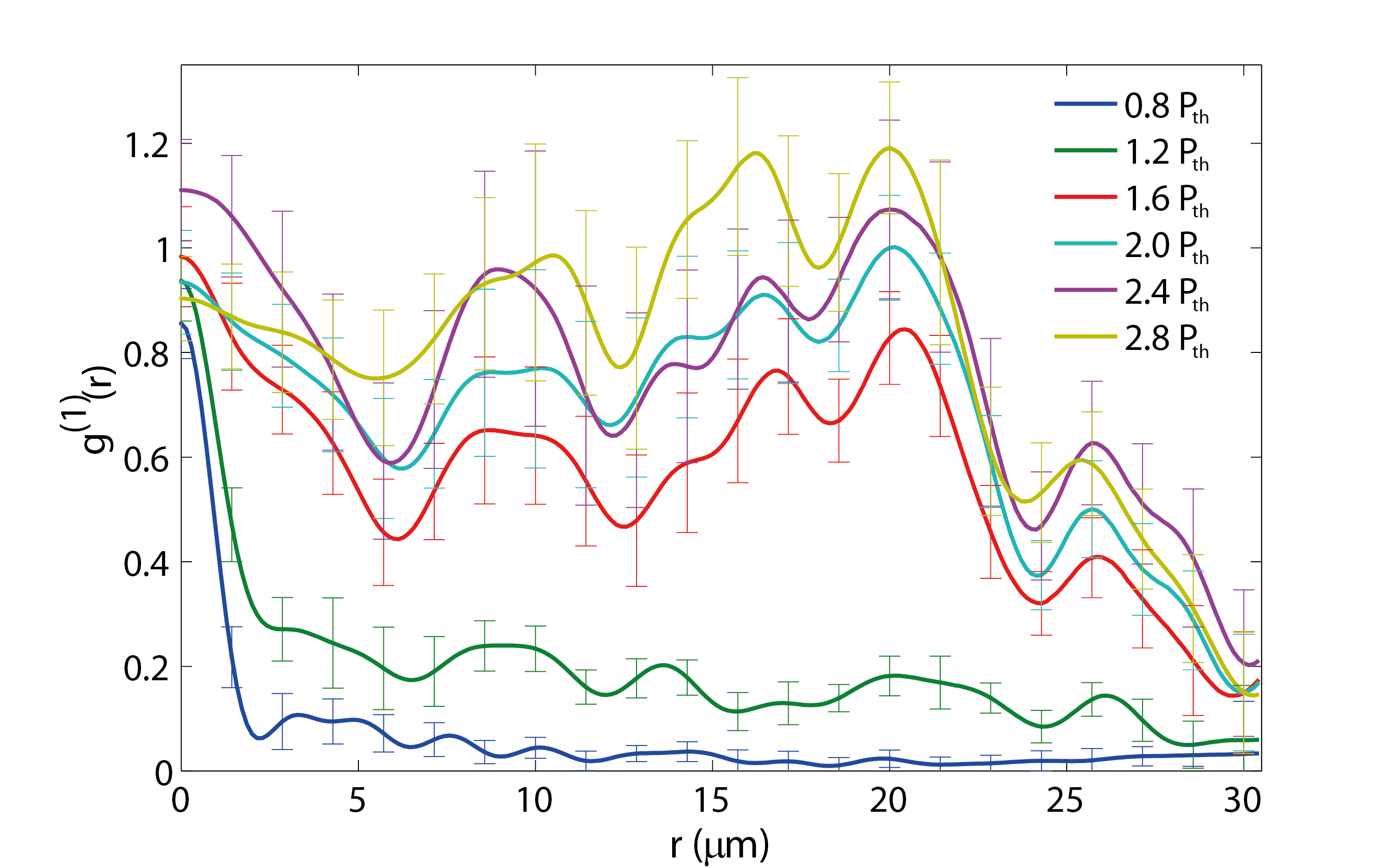}
\caption{\label{fig:figure2} First-order spatial coherence function along the linear condensate for different values of the pump power. The threshold power $P_{th}$ for condensation is found to be around 100 $\mu$W. The spatial modulation of the $g^{(1)}$ reflects the density modulation due to the presence of disorder (see text). The error bars quantify the uncertainty in the numerical determination of the $g^{(1)}$ value.}
\end{figure}

In Fig.\ref{fig:figure2} we show the $g^{(1)}(r)$ for different excitation powers, $r = 0$ being the autocorrelation point. Given the centrosymmetry of the interferogram, we plot curves that result from the average of the two spatial halves of the coherence function.
Below threshold the correlation is limited to a small region around the autocorrelation point, whose size is approximately 0.8 $\mu$m, defined by the convolution of the response function of the microscope objective and the thermal de Broglie wavelength $\lambda_{th}$ of polaritons \cite{kasprzak_bose-einstein_2006}. As the threshold power is reached, coherence gradually builds up along the 1D disorder valley. Further increasing the power, long-range order establishes and the coherence extends over the whole length of the 1D condensate.

A striking feature in Fig.\ref{fig:figure2} is the presence of a spatial modulation of the $g^{(1)}(r)$. This effect can be explained taking into account the condensate density, shown in Fig.\ref{fig:figure3}.a.  The expression for the $g^{(1)}$ as a function of the condensate density, $n_{ic}$, and the associated density fluctuations, $\delta n_{i}$, reads
\begin{equation}\label{eq:g1mod}
g^{(1)} = \frac{\sqrt{n_{1c}n_{2c}}}{\sqrt{(n_{1c} + \delta n_{1})(n_{2c} + \delta n_{2})}} = \frac{1}{\sqrt{(1 + \frac{\delta n_{1}}{n_{1c}})(1 + \frac{\delta n_{2}}{n_{2c}})}}
\end{equation}
where $i=1,2$ indicates the two arms of the interferometer. If we assume that the incoherent density $\delta n_i$ is independent of the potential fluctuations, it is immediately clear that the minima of the spatial coherence and total density coincide.

We want to point our that the present coherence measurements yield the largest values of the long range order measured up to now in CdTe microcavities. The previously measured values of the long-range spatial coherence were less than 0.5. These lower values are due to the formation of multiple condensates on generic positions on the sample.

In conclusion, in this work we have presented a study of the coherence properties of a non-equilibrium exciton-polariton condensate in a one-dimensional potential trap. We have reported on the behavior of the first-order coherence function with increasing pump power, highlighting the relation between its spatial modulation and the modulation of the 1D condensate density due to the presence of a disorder potential. As in one-dimension the effect of disorder in more pronounced, we are able to experimentally measure a non-monotonous coherence function, resulting from the strong non-equilibrium nature of our polariton system.

In a view of the identification and possible experimental assessment of the BG phase, further experiments on new disorder-engineered samples should be performed, in order to be able to design and control the disorder landscape and amplitude felt by the polariton quantum fluid.

This work was supported by the Swiss National Science Foundation through NCCR ``Quantum Photonics''.

\bibliography{paper1D}

\providecommand{\noopsort}[1]{}\providecommand{\singleletter}[1]{#1}%
\begin{thebibliography}{25}
\expandafter\ifx\csname natexlab\endcsname\relax\def\natexlab#1{#1}\fi
\expandafter\ifx\csname bibnamefont\endcsname\relax
  \def\bibnamefont#1{#1}\fi
\expandafter\ifx\csname bibfnamefont\endcsname\relax
  \def\bibfnamefont#1{#1}\fi
\expandafter\ifx\csname citenamefont\endcsname\relax
  \def\citenamefont#1{#1}\fi
\expandafter\ifx\csname url\endcsname\relax
  \def\url#1{\texttt{#1}}\fi
\expandafter\ifx\csname urlprefix\endcsname\relax\def\urlprefix{URL }\fi
\providecommand{\bibinfo}[2]{#2}
\providecommand{\eprint}[2][]{\url{#2}}

\bibitem[{\citenamefont{Reppy}(1992)}]{reppy_superfluid_1992}
\bibinfo{author}{\bibfnamefont{J.~D.} \bibnamefont{Reppy}},
  \bibinfo{journal}{Journal of Low Temperature Physics}
  \textbf{\bibinfo{volume}{87}}, \bibinfo{pages}{205} (\bibinfo{year}{1992}).

\bibitem[{\citenamefont{Pitaevskii and Stringari}(2003)}]{stringari_bec}
\bibinfo{author}{\bibfnamefont{L.~P.} \bibnamefont{Pitaevskii}}
  \bibnamefont{and}
  \bibinfo{author}{\bibfnamefont{S.}~\bibnamefont{Stringari}},
  \emph{\bibinfo{title}{Bose-Einstein Condensation}}
  (\bibinfo{publisher}{Clarendon Press, Oxford}, \bibinfo{year}{2003}).

\bibitem[{\citenamefont{{Sanchez-Palencia} and
  Lewenstein}(2010)}]{sanchez-palencia_disordered_2010}
\bibinfo{author}{\bibfnamefont{L.}~\bibnamefont{{Sanchez-Palencia}}}
  \bibnamefont{and}
  \bibinfo{author}{\bibfnamefont{M.}~\bibnamefont{Lewenstein}},
  \bibinfo{journal}{Nat Phys} \textbf{\bibinfo{volume}{6}}, \bibinfo{pages}{87}
  (\bibinfo{year}{2010}), ISSN \bibinfo{issn}{1745-2473}.

\bibitem[{\citenamefont{Greiner et~al.}(2002)\citenamefont{Greiner, Mandel,
  Esslinger, Hansch, and Bloch}}]{greiner_quantum_2002}
\bibinfo{author}{\bibfnamefont{M.}~\bibnamefont{Greiner}},
  \bibinfo{author}{\bibfnamefont{O.}~\bibnamefont{Mandel}},
  \bibinfo{author}{\bibfnamefont{T.}~\bibnamefont{Esslinger}},
  \bibinfo{author}{\bibfnamefont{T.~W.} \bibnamefont{Hansch}},
  \bibnamefont{and} \bibinfo{author}{\bibfnamefont{I.}~\bibnamefont{Bloch}},
  \bibinfo{journal}{Nature} \textbf{\bibinfo{volume}{415}}, \bibinfo{pages}{39}
  (\bibinfo{year}{2002}), ISSN \bibinfo{issn}{0028-0836}.

\bibitem[{\citenamefont{Deissler et~al.}(2010)\citenamefont{Deissler, Zaccanti,
  Roati, {D'Errico}, Fattori, Modugno, Modugno, and
  Inguscio}}]{deissler_delocalization_2010}
\bibinfo{author}{\bibfnamefont{B.}~\bibnamefont{Deissler}},
  \bibinfo{author}{\bibfnamefont{M.}~\bibnamefont{Zaccanti}},
  \bibinfo{author}{\bibfnamefont{G.}~\bibnamefont{Roati}},
  \bibinfo{author}{\bibfnamefont{C.}~\bibnamefont{{D'Errico}}},
  \bibinfo{author}{\bibfnamefont{M.}~\bibnamefont{Fattori}},
  \bibinfo{author}{\bibfnamefont{M.}~\bibnamefont{Modugno}},
  \bibinfo{author}{\bibfnamefont{G.}~\bibnamefont{Modugno}}, \bibnamefont{and}
  \bibinfo{author}{\bibfnamefont{M.}~\bibnamefont{Inguscio}},
  \bibinfo{journal}{Nat Phys} \textbf{\bibinfo{volume}{6}},
  \bibinfo{pages}{354} (\bibinfo{year}{2010}), ISSN \bibinfo{issn}{1745-2473}.

\bibitem[{\citenamefont{Pasienski et~al.}(2010)\citenamefont{Pasienski,
  {McKay}, White, and {DeMarco}}}]{pasienski_disordered_2010}
\bibinfo{author}{\bibfnamefont{M.}~\bibnamefont{Pasienski}},
  \bibinfo{author}{\bibfnamefont{D.}~\bibnamefont{{McKay}}},
  \bibinfo{author}{\bibfnamefont{M.}~\bibnamefont{White}}, \bibnamefont{and}
  \bibinfo{author}{\bibfnamefont{B.}~\bibnamefont{{DeMarco}}},
  \bibinfo{journal}{Nat Phys} \textbf{\bibinfo{volume}{6}},
  \bibinfo{pages}{677} (\bibinfo{year}{2010}), ISSN \bibinfo{issn}{1745-2473}.

\bibitem[{\citenamefont{Fisher et~al.}(1989)\citenamefont{Fisher, Weichman,
  Grinstein, and Fisher}}]{fisher_boson_1989}
\bibinfo{author}{\bibfnamefont{M.~P.~A.} \bibnamefont{Fisher}},
  \bibinfo{author}{\bibfnamefont{P.~B.} \bibnamefont{Weichman}},
  \bibinfo{author}{\bibfnamefont{G.}~\bibnamefont{Grinstein}},
  \bibnamefont{and} \bibinfo{author}{\bibfnamefont{D.~S.}
  \bibnamefont{Fisher}}, \bibinfo{journal}{Physical Review B}
  \textbf{\bibinfo{volume}{40}}, \bibinfo{pages}{546} (\bibinfo{year}{1989}).

\bibitem[{\citenamefont{Chen et~al.}(2008)\citenamefont{Chen, Hitchcock, Dries,
  Junker, Welford, and Hulet}}]{chen_phase_2008}
\bibinfo{author}{\bibfnamefont{Y.~P.} \bibnamefont{Chen}},
  \bibinfo{author}{\bibfnamefont{J.}~\bibnamefont{Hitchcock}},
  \bibinfo{author}{\bibfnamefont{D.}~\bibnamefont{Dries}},
  \bibinfo{author}{\bibfnamefont{M.}~\bibnamefont{Junker}},
  \bibinfo{author}{\bibfnamefont{C.}~\bibnamefont{Welford}}, \bibnamefont{and}
  \bibinfo{author}{\bibfnamefont{R.~G.} \bibnamefont{Hulet}},
  \bibinfo{journal}{Physical Review A} \textbf{\bibinfo{volume}{77}},
  \bibinfo{pages}{033632} (\bibinfo{year}{2008}).

\bibitem[{\citenamefont{Roati et~al.}(2008)\citenamefont{Roati, {D/'Errico},
  Fallani, Fattori, Fort, Zaccanti, Modugno, Modugno, and
  Inguscio}}]{roati_anderson_2008}
\bibinfo{author}{\bibfnamefont{G.}~\bibnamefont{Roati}},
  \bibinfo{author}{\bibfnamefont{C.}~\bibnamefont{{D/'Errico}}},
  \bibinfo{author}{\bibfnamefont{L.}~\bibnamefont{Fallani}},
  \bibinfo{author}{\bibfnamefont{M.}~\bibnamefont{Fattori}},
  \bibinfo{author}{\bibfnamefont{C.}~\bibnamefont{Fort}},
  \bibinfo{author}{\bibfnamefont{M.}~\bibnamefont{Zaccanti}},
  \bibinfo{author}{\bibfnamefont{G.}~\bibnamefont{Modugno}},
  \bibinfo{author}{\bibfnamefont{M.}~\bibnamefont{Modugno}}, \bibnamefont{and}
  \bibinfo{author}{\bibfnamefont{M.}~\bibnamefont{Inguscio}},
  \bibinfo{journal}{Nature} \textbf{\bibinfo{volume}{453}},
  \bibinfo{pages}{895} (\bibinfo{year}{2008}), ISSN \bibinfo{issn}{0028-0836}.

\bibitem[{\citenamefont{Billy et~al.}(2008)\citenamefont{Billy, Josse, Zuo,
  Bernard, Hambrecht, Lugan, Clement, {Sanchez-Palencia}, Bouyer, and
  Aspect}}]{billy_direct_2008}
\bibinfo{author}{\bibfnamefont{J.}~\bibnamefont{Billy}},
  \bibinfo{author}{\bibfnamefont{V.}~\bibnamefont{Josse}},
  \bibinfo{author}{\bibfnamefont{Z.}~\bibnamefont{Zuo}},
  \bibinfo{author}{\bibfnamefont{A.}~\bibnamefont{Bernard}},
  \bibinfo{author}{\bibfnamefont{B.}~\bibnamefont{Hambrecht}},
  \bibinfo{author}{\bibfnamefont{P.}~\bibnamefont{Lugan}},
  \bibinfo{author}{\bibfnamefont{D.}~\bibnamefont{Clement}},
  \bibinfo{author}{\bibfnamefont{L.}~\bibnamefont{{Sanchez-Palencia}}},
  \bibinfo{author}{\bibfnamefont{P.}~\bibnamefont{Bouyer}}, \bibnamefont{and}
  \bibinfo{author}{\bibfnamefont{A.}~\bibnamefont{Aspect}},
  \bibinfo{journal}{Nature} \textbf{\bibinfo{volume}{453}},
  \bibinfo{pages}{891} (\bibinfo{year}{2008}), ISSN \bibinfo{issn}{0028-0836}.

\bibitem[{\citenamefont{Kasprzak et~al.}(2006)\citenamefont{Kasprzak, Richard,
  Kundermann, Baas, Jeambrun, Keeling, Marchetti, Szymanska, Andre, Staehli
  et~al.}}]{kasprzak_bose-einstein_2006}
\bibinfo{author}{\bibfnamefont{J.}~\bibnamefont{Kasprzak}},
  \bibinfo{author}{\bibfnamefont{M.}~\bibnamefont{Richard}},
  \bibinfo{author}{\bibfnamefont{S.}~\bibnamefont{Kundermann}},
  \bibinfo{author}{\bibfnamefont{A.}~\bibnamefont{Baas}},
  \bibinfo{author}{\bibfnamefont{P.}~\bibnamefont{Jeambrun}},
  \bibinfo{author}{\bibfnamefont{J.~M.~J.} \bibnamefont{Keeling}},
  \bibinfo{author}{\bibfnamefont{F.~M.} \bibnamefont{Marchetti}},
  \bibinfo{author}{\bibfnamefont{M.~H.} \bibnamefont{Szymanska}},
  \bibinfo{author}{\bibfnamefont{R.}~\bibnamefont{Andre}},
  \bibinfo{author}{\bibfnamefont{J.~L.} \bibnamefont{Staehli}},
  \bibnamefont{et~al.}, \bibinfo{journal}{Nature}
  \textbf{\bibinfo{volume}{443}}, \bibinfo{pages}{409} (\bibinfo{year}{2006}),
  ISSN \bibinfo{issn}{0028-0836}.

\bibitem[{\citenamefont{Balili et~al.}(2007)\citenamefont{Balili, Hartwell,
  Snoke, Pfeiffer, and West}}]{balili_bose-einstein_2007}
\bibinfo{author}{\bibfnamefont{R.}~\bibnamefont{Balili}},
  \bibinfo{author}{\bibfnamefont{V.}~\bibnamefont{Hartwell}},
  \bibinfo{author}{\bibfnamefont{D.}~\bibnamefont{Snoke}},
  \bibinfo{author}{\bibfnamefont{L.}~\bibnamefont{Pfeiffer}}, \bibnamefont{and}
  \bibinfo{author}{\bibfnamefont{K.}~\bibnamefont{West}},
  \bibinfo{journal}{Science} \textbf{\bibinfo{volume}{316}},
  \bibinfo{pages}{1007} (\bibinfo{year}{2007}).

\bibitem[{\citenamefont{Deng et~al.}(2007)\citenamefont{Deng, Solomon, Hey,
  Ploog, and Yamamoto}}]{deng_spatial_2007}
\bibinfo{author}{\bibfnamefont{H.}~\bibnamefont{Deng}},
  \bibinfo{author}{\bibfnamefont{G.~S.} \bibnamefont{Solomon}},
  \bibinfo{author}{\bibfnamefont{R.}~\bibnamefont{Hey}},
  \bibinfo{author}{\bibfnamefont{K.~H.} \bibnamefont{Ploog}}, \bibnamefont{and}
  \bibinfo{author}{\bibfnamefont{Y.}~\bibnamefont{Yamamoto}},
  \bibinfo{journal}{Physical Review Letters} \textbf{\bibinfo{volume}{99}},
  \bibinfo{pages}{126403} (\bibinfo{year}{2007}).

\bibitem[{\citenamefont{Wertz et~al.}(2010)\citenamefont{Wertz, Ferrier,
  Solnyshkov, Johne, Sanvitto, Lemaitre, Sagnes, Grousson, Kavokin, Senellart
  et~al.}}]{wertz_spontaneous_2010}
\bibinfo{author}{\bibfnamefont{E.}~\bibnamefont{Wertz}},
  \bibinfo{author}{\bibfnamefont{L.}~\bibnamefont{Ferrier}},
  \bibinfo{author}{\bibfnamefont{D.~D.} \bibnamefont{Solnyshkov}},
  \bibinfo{author}{\bibfnamefont{R.}~\bibnamefont{Johne}},
  \bibinfo{author}{\bibfnamefont{D.}~\bibnamefont{Sanvitto}},
  \bibinfo{author}{\bibfnamefont{A.}~\bibnamefont{Lemaitre}},
  \bibinfo{author}{\bibfnamefont{I.}~\bibnamefont{Sagnes}},
  \bibinfo{author}{\bibfnamefont{R.}~\bibnamefont{Grousson}},
  \bibinfo{author}{\bibfnamefont{A.~V.} \bibnamefont{Kavokin}},
  \bibinfo{author}{\bibfnamefont{P.}~\bibnamefont{Senellart}},
  \bibnamefont{et~al.}, \bibinfo{journal}{Nat Phys}
  \textbf{\bibinfo{volume}{6}}, \bibinfo{pages}{860} (\bibinfo{year}{2010}),
  ISSN \bibinfo{issn}{1745-2473}.

\bibitem[{\citenamefont{Christmann et~al.}(2008)\citenamefont{Christmann,
  ButteÌ, Feltin, Carlin, and Grandjean}}]{christmann_room_2008}
\bibinfo{author}{\bibfnamefont{G.}~\bibnamefont{Christmann}},
  \bibinfo{author}{\bibfnamefont{R.}~\bibnamefont{ButteÌ}},
  \bibinfo{author}{\bibfnamefont{E.}~\bibnamefont{Feltin}},
  \bibinfo{author}{\bibfnamefont{J.}~\bibnamefont{Carlin}}, \bibnamefont{and}
  \bibinfo{author}{\bibfnamefont{N.}~\bibnamefont{Grandjean}},
  \bibinfo{journal}{Applied Physics Letters} \textbf{\bibinfo{volume}{93}},
  \bibinfo{pages}{051102} (\bibinfo{year}{2008}), ISSN
  \bibinfo{issn}{00036951}.

\bibitem[{\citenamefont{Lagoudakis et~al.}(2008)\citenamefont{Lagoudakis,
  Wouters, Richard, Baas, Carusotto, Andre, Dang, and
  {Deveaud-Pl\'edran}}}]{lagoudakis_quantized_2008}
\bibinfo{author}{\bibfnamefont{K.~G.} \bibnamefont{Lagoudakis}},
  \bibinfo{author}{\bibfnamefont{M.}~\bibnamefont{Wouters}},
  \bibinfo{author}{\bibfnamefont{M.}~\bibnamefont{Richard}},
  \bibinfo{author}{\bibfnamefont{A.}~\bibnamefont{Baas}},
  \bibinfo{author}{\bibfnamefont{I.}~\bibnamefont{Carusotto}},
  \bibinfo{author}{\bibfnamefont{R.}~\bibnamefont{Andre}},
  \bibinfo{author}{\bibfnamefont{L.~S.} \bibnamefont{Dang}}, \bibnamefont{and}
  \bibinfo{author}{\bibfnamefont{B.}~\bibnamefont{{Deveaud-Pl\'edran}}},
  \bibinfo{journal}{Nat Phys} \textbf{\bibinfo{volume}{4}},
  \bibinfo{pages}{706} (\bibinfo{year}{2008}), ISSN \bibinfo{issn}{1745-2473}.

\bibitem[{\citenamefont{Lagoudakis et~al.}(2009)\citenamefont{Lagoudakis,
  Ostatnicky, Kavokin, Rubo, Andre, and
  {Deveaud-Pl\'edran}}}]{lagoudakis_observation_2009}
\bibinfo{author}{\bibfnamefont{K.~G.} \bibnamefont{Lagoudakis}},
  \bibinfo{author}{\bibfnamefont{T.}~\bibnamefont{Ostatnicky}},
  \bibinfo{author}{\bibfnamefont{A.~V.} \bibnamefont{Kavokin}},
  \bibinfo{author}{\bibfnamefont{Y.~G.} \bibnamefont{Rubo}},
  \bibinfo{author}{\bibfnamefont{R.}~\bibnamefont{Andre}}, \bibnamefont{and}
  \bibinfo{author}{\bibfnamefont{B.}~\bibnamefont{{Deveaud-Pl\'edran}}},
  \bibinfo{journal}{Science} \textbf{\bibinfo{volume}{326}},
  \bibinfo{pages}{974} (\bibinfo{year}{2009}).

\bibitem[{\citenamefont{Amo et~al.}(2009{\natexlab{a}})\citenamefont{Amo,
  Sanvitto, Laussy, Ballarini, del Valle, Martin, Lemaitre, Bloch,
  Krizhanovskii, Skolnick et~al.}}]{amo_collective_2009}
\bibinfo{author}{\bibfnamefont{A.}~\bibnamefont{Amo}},
  \bibinfo{author}{\bibfnamefont{D.}~\bibnamefont{Sanvitto}},
  \bibinfo{author}{\bibfnamefont{F.~P.} \bibnamefont{Laussy}},
  \bibinfo{author}{\bibfnamefont{D.}~\bibnamefont{Ballarini}},
  \bibinfo{author}{\bibfnamefont{E.}~\bibnamefont{del Valle}},
  \bibinfo{author}{\bibfnamefont{M.~D.} \bibnamefont{Martin}},
  \bibinfo{author}{\bibfnamefont{A.}~\bibnamefont{Lemaitre}},
  \bibinfo{author}{\bibfnamefont{J.}~\bibnamefont{Bloch}},
  \bibinfo{author}{\bibfnamefont{D.~N.} \bibnamefont{Krizhanovskii}},
  \bibinfo{author}{\bibfnamefont{M.~S.} \bibnamefont{Skolnick}},
  \bibnamefont{et~al.}, \bibinfo{journal}{Nature}
  \textbf{\bibinfo{volume}{457}}, \bibinfo{pages}{291}
  (\bibinfo{year}{2009}{\natexlab{a}}), ISSN \bibinfo{issn}{0028-0836}.

\bibitem[{\citenamefont{Amo et~al.}(2009{\natexlab{b}})\citenamefont{Amo,
  Lefrere, Pigeon, Adrados, Ciuti, Carusotto, Houdre, Giacobino, and
  Bramati}}]{amo_superfluidity_2009}
\bibinfo{author}{\bibfnamefont{A.}~\bibnamefont{Amo}},
  \bibinfo{author}{\bibfnamefont{J.}~\bibnamefont{Lefrere}},
  \bibinfo{author}{\bibfnamefont{S.}~\bibnamefont{Pigeon}},
  \bibinfo{author}{\bibfnamefont{C.}~\bibnamefont{Adrados}},
  \bibinfo{author}{\bibfnamefont{C.}~\bibnamefont{Ciuti}},
  \bibinfo{author}{\bibfnamefont{I.}~\bibnamefont{Carusotto}},
  \bibinfo{author}{\bibfnamefont{R.}~\bibnamefont{Houdre}},
  \bibinfo{author}{\bibfnamefont{E.}~\bibnamefont{Giacobino}},
  \bibnamefont{and} \bibinfo{author}{\bibfnamefont{A.}~\bibnamefont{Bramati}},
  \bibinfo{journal}{Nat Phys} \textbf{\bibinfo{volume}{5}},
  \bibinfo{pages}{805} (\bibinfo{year}{2009}{\natexlab{b}}), ISSN
  \bibinfo{issn}{1745-2473}.

\bibitem[{\citenamefont{Daïf et~al.}(2006)\citenamefont{Daïf, Baas, Guillet,
  Brantut, Kaitouni, Staehli, {Morier-Genoud}, and
  {Deveaud-Pl\'edran}}}]{el_daif_polariton_2006}
\bibinfo{author}{\bibfnamefont{O.~E.} \bibnamefont{Daïf}},
  \bibinfo{author}{\bibfnamefont{A.}~\bibnamefont{Baas}},
  \bibinfo{author}{\bibfnamefont{T.}~\bibnamefont{Guillet}},
  \bibinfo{author}{\bibfnamefont{J.}~\bibnamefont{Brantut}},
  \bibinfo{author}{\bibfnamefont{R.~I.} \bibnamefont{Kaitouni}},
  \bibinfo{author}{\bibfnamefont{J.~L.} \bibnamefont{Staehli}},
  \bibinfo{author}{\bibfnamefont{F.}~\bibnamefont{{Morier-Genoud}}},
  \bibnamefont{and}
  \bibinfo{author}{\bibfnamefont{B.}~\bibnamefont{{Deveaud-Pl\'edran}}},
  \bibinfo{journal}{Applied Physics Letters} \textbf{\bibinfo{volume}{88}},
  \bibinfo{pages}{061105} (\bibinfo{year}{2006}), ISSN
  \bibinfo{issn}{00036951}.

\bibitem[{\citenamefont{Wouters et~al.}(2010)\citenamefont{Wouters, Liew, and
  Savona}}]{wouters_energy_2010}
\bibinfo{author}{\bibfnamefont{M.}~\bibnamefont{Wouters}},
  \bibinfo{author}{\bibfnamefont{T.~C.~H.} \bibnamefont{Liew}},
  \bibnamefont{and} \bibinfo{author}{\bibfnamefont{V.}~\bibnamefont{Savona}},
  \bibinfo{journal}{Physical Review B} \textbf{\bibinfo{volume}{82}},
  \bibinfo{pages}{245315} (\bibinfo{year}{2010}).

\bibitem[{\citenamefont{Krizhanovskii et~al.}(2009)\citenamefont{Krizhanovskii,
  Lagoudakis, Wouters, Pietka, Bradley, Guda, Whittaker, Skolnick,
  {Deveaud-Pl\'edran}, Richard et~al.}}]{krizhanovskii_coexisting_2009}
\bibinfo{author}{\bibfnamefont{D.~N.} \bibnamefont{Krizhanovskii}},
  \bibinfo{author}{\bibfnamefont{K.~G.} \bibnamefont{Lagoudakis}},
  \bibinfo{author}{\bibfnamefont{M.}~\bibnamefont{Wouters}},
  \bibinfo{author}{\bibfnamefont{B.}~\bibnamefont{Pietka}},
  \bibinfo{author}{\bibfnamefont{R.~A.} \bibnamefont{Bradley}},
  \bibinfo{author}{\bibfnamefont{K.}~\bibnamefont{Guda}},
  \bibinfo{author}{\bibfnamefont{D.~M.} \bibnamefont{Whittaker}},
  \bibinfo{author}{\bibfnamefont{M.~S.} \bibnamefont{Skolnick}},
  \bibinfo{author}{\bibfnamefont{B.}~\bibnamefont{{Deveaud-Pl\'edran}}},
  \bibinfo{author}{\bibfnamefont{M.}~\bibnamefont{Richard}},
  \bibnamefont{et~al.}, \bibinfo{journal}{Physical Review B}
  \textbf{\bibinfo{volume}{80}}, \bibinfo{pages}{045317}
  (\bibinfo{year}{2009}).

\bibitem[{\citenamefont{Baas et~al.}(2008)\citenamefont{Baas, Lagoudakis,
  Richard, Andrï¿½, Dang, and {Deveaud-Pl\'edran}}}]{baas_synchronized_2008}
\bibinfo{author}{\bibfnamefont{A.}~\bibnamefont{Baas}},
  \bibinfo{author}{\bibfnamefont{K.~G.} \bibnamefont{Lagoudakis}},
  \bibinfo{author}{\bibfnamefont{M.}~\bibnamefont{Richard}},
  \bibinfo{author}{\bibfnamefont{R.}~\bibnamefont{Andrï¿½}},
  \bibinfo{author}{\bibfnamefont{L.~S.} \bibnamefont{Dang}}, \bibnamefont{and}
  \bibinfo{author}{\bibfnamefont{B.}~\bibnamefont{{Deveaud-Pl\'edran}}},
  \bibinfo{journal}{Physical Review Letters} \textbf{\bibinfo{volume}{100}},
  \bibinfo{pages}{170401} (\bibinfo{year}{2008}).

\bibitem[{\citenamefont{Cl\'ement et~al.}(2008)\citenamefont{Cl\'ement, Bouyer,
  Aspect, and {Sanchez-Palencia}}}]{clement_density_2008}
\bibinfo{author}{\bibfnamefont{D.}~\bibnamefont{Cl\'ement}},
  \bibinfo{author}{\bibfnamefont{P.}~\bibnamefont{Bouyer}},
  \bibinfo{author}{\bibfnamefont{A.}~\bibnamefont{Aspect}}, \bibnamefont{and}
  \bibinfo{author}{\bibfnamefont{L.}~\bibnamefont{{Sanchez-Palencia}}},
  \bibinfo{journal}{Physical Review A} \textbf{\bibinfo{volume}{77}},
  \bibinfo{pages}{033631} (\bibinfo{year}{2008}).

\bibitem[{ref()}]{ref_prof_details}
\bibinfo{note}{Other kinds of smooth profiles, like top-hat and combined
  gaussian-top hat have been tested as reference profile for our analysis, but
  the results were found to be not substantially sensitive to the assumed
  reference profile to the aim of our analysis.}

\end{thebibliography}

\end{document}